\documentclass[8.5pt,twoside,twocolumn]{article}
\usepackage[super,sort&compress,comma]{natbib}
\usepackage{mhchem}
\usepackage{color,soul}
\usepackage{times}
\usepackage{sectsty}
\usepackage{balance}

\usepackage{graphicx} 
\usepackage{lastpage}
\usepackage[format=plain,justification=raggedright,singlelinecheck=false,font=small,labelfont=bf,labelsep=space]{caption}
\usepackage{fancyhdr}

\usepackage[normalem]{ulem}
\usepackage{amssymb}

\usepackage{multicol,caption}

\begin{document}

\thispagestyle{plain}
\fancypagestyle{plain}{
\fancyhead[L]{\includegraphics[height=8pt]{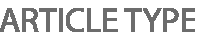}}
\fancyhead[C]{\hspace{-1cm}\includegraphics[height=20pt]{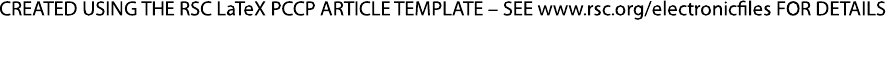}}
\fancyhead[R]{\includegraphics[height=10pt]{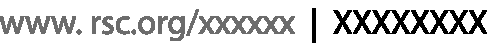}\vspace{-0.2cm}}
\renewcommand{\headrulewidth}{1pt}}
\renewcommand{\thefootnote}{\fnsymbol{footnote}}
\renewcommand\footnoterule{\vspace*{1pt}%
\hrule width 3.4in height 0.4pt \vspace*{5pt}}
\setcounter{secnumdepth}{5}

\makeatletter
\def\subsubsection{\@startsection{subsubsection}{3}{10pt}{-1.25ex plus -1ex minus -.1ex}{0ex plus 0ex}{\normalsize\bf}}
\def\paragraph{\@startsection{paragraph}{4}{10pt}{-1.25ex plus -1ex minus -.1ex}{0ex plus 0ex}{\normalsize\textit}}
\renewcommand\@biblabel[1]{#1}
\renewcommand\@makefntext[1]%
{\noindent\makebox[0pt][r]{\@thefnmark\,}#1}
\makeatother
\renewcommand{\figurename}{\small{Fig.}~}
\sectionfont{\large}
\subsectionfont{\normalsize}

\fancyfoot{}
\fancyfoot[LO,RE]{\vspace{-7pt}\includegraphics[height=9pt]{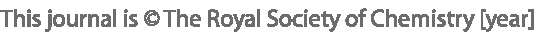}}
\fancyfoot[CO]{\vspace{-7.2pt}\hspace{12.2cm}\includegraphics{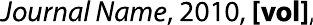}}
\fancyfoot[CE]{\vspace{-7.5pt}\hspace{-13.5cm}\includegraphics{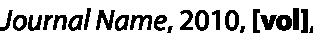}}
\fancyfoot[RO]{\footnotesize{\sffamily{1--\pageref{LastPage} ~\textbar  \hspace{2pt}\thepage}}}
\fancyfoot[LE]{\footnotesize{\sffamily{\thepage~\textbar\hspace{3.45cm} 1--\pageref{LastPage}}}}
\fancyhead{}
\renewcommand{\headrulewidth}{1pt}
\renewcommand{\footrulewidth}{1pt}
\setlength{\arrayrulewidth}{1pt}
\setlength{\columnsep}{6.5mm}
\setlength\bibsep{1pt}

\twocolumn[
  \begin{@twocolumnfalse}
\noindent\LARGE{\textbf{Single file dynamics in soft materials}}
\vspace{0.6cm}

\noindent\large{\textbf{Alessandro Taloni,\textit{$^{a\ddag}$} Ophir Flomenbom,\textit{$^{b\ddag}$} Ram\'on Casta\~neda-Priego,\textit{$^{c}$} Fabio Marchesoni,\textit{$^{d}$} }}
\vspace{0.5cm}

\noindent\textit{\small{\textbf{Received Xth XXXXXXXXXX 20XX, Accepted Xth XXXXXXXXX 20XX\newline
First published on the web Xth XXXXXXXXXX 200X}}}

\noindent \textbf{\small{DOI: 10.1039/b000000x}}
\vspace{0.6cm}

\noindent \normalsize{The term single file (SF) dynamics refers to the motion of an assembly of particles through a channel with cross-section comparable to the particles' diameter. Single file diffusion (SFD) is then the diffusion of a tagged particle in a \emph{single file}, i.e., under the condition that particle passing is not allowed.  SFD accounts for a large variety of processes in nature, including diffusion of colloids in synthetic and natural-shaped channels,  biological motors along molecular chains, electrons in proteins and liquid helium, ions through membrane, just to mention a few examples. Albeit introduced in '65, over the last decade the classical notion of SF dynamics has been generalized to account through a more realistic modeling of, among others, particles properties, file geometry, particle-particle and channel-particles interactions, thus paving the way to remarkable applications in, for example, the technology of bio-integrated nanodevices. We then provide a comprehensive review of the recent advances in the theory of SF dynamics and the ensuing experimental realisations.}
\vspace{0.5cm}
 \end{@twocolumnfalse}
  ]

\footnotetext{\textit{$^{a}$~Center for Complexity \& Biosystems, Physics Department, University of Milan ``La Statale'', Via Giovanni Celoria, 16, 20133  Milano, Italy; E-mail: alessandro.taloni@gmail.com}}
\footnotetext{\textit{$^{b}$~Flomenbom-BPS Ltd, 19 Louis Marshal St, Tel Aviv, Israel. E-mail: ophir1974@flomenbom.net
 }}
\footnotetext{\textit{$^{c}$Division of Science and Engineering, University of Guanajuato, Loma del Bosque 103, Lomas del Campestre, 37150, Leon, Gto., Mexico. E-mail: ramoncp@fisica.ugto.mx}}

\footnotetext{\textit{$^{d}$~Dipartimento di Fisica, Universit\'a di Camerino, I-62032 Camerino, Italy. fabio.marchesoni@unicam.it}}

\footnotetext{\ddag~These authors contributed equally to this work.}

\section{Introduction}


The single-file (SF) concept was introduced first by physiologists in 1954.
In their experiment, Hodgkin and Keynes \cite{hodgkin1955} resorted to the notion of no-passing particles in a pore, to account for the observed anomalous decay of the influx of potassium ions moving inwards and outwards across the membranes of giant axons of {\it Sepia officinalis}.
Implicit in their analysis, and further elaborated on by E. J. Harris\cite{harris1960}, Lea\cite{lea1963}, Rickert\cite{rickert1964} and Heckmann\cite{heckmann1972}, was the idea that the osmotic permeability coefficient through the membrane pores is larger than the diffusive permeability, and their ratio equals the number of molecules in the channel.
This result was stated explicitly in the works of Dick\cite{dick1966} and Levitt\cite{levitt1974}, and experimentally demonstrated by Rosenberg and Finkelstein\cite{Rosenberg1978}, who used it to estimate the number of molecules inside the pores.
The central problem addressed by these authors was the tracer diffusion, i.e., the diffusion coefficient of a single distinguishable particle in a file moving through pores of finite length and and diameter close to that of the translocating molecules: At odds with common wisdom, they noticed that the tracer diffusion coefficient went to zero as the length of the pore (or the membrane thickness) increased.

During the Sixties, the problem of molecular diffusion in narrow pores was considered an interesting though somewhat exotic topic and as such was investigated at depth by the mathematical-physical community.
The SF model was formulated in its present form by T. Harris in 1965\cite{Harris_JApplProb_1965}, who generalized the SF condition to thermodynamic systems, i.e., for infinitely long pores and infinitely many no-passing diffusing particles, but constant linear number density, $\rho$. The particles were modeled as identical hard-core  point-like particles subject to Brownian motion on an infinite line. The SF condition was implemented by requiring that energy and momentum be conserved at each collision. It follows immediately that a tagged particle (or tracer) undergoes asymptotic subdiffusive behaviour with law,
\begin{equation}
\langle \left[x(t)-x(0)\right]^2\rangle=2\sqrt{\frac{Dt}{\pi\rho^2}}.
\label{SFD}
 \end{equation}
with Gaussian probability density function (PDF) -- a result rigorously proven few years later by Arratia \cite{Arratia_AnnProb_1983}). Here, $x(t)$ represents the tagged particle coordinate along the 1D substrate, $D$ is the diffusion coefficient of a free diffusing particle in the bulk (i.e., away from the pore walls and the other file particles), and $\langle\cdots\rangle$ denotes the ensemble average. Since its formal introduction, the single file diffusion (SFD) law of Eq. (\ref{SFD}) has attracted the attention of a growing number of investigators from the most diverse scientific backgrounds for two main reasons. First, because it is perhaps the simplest analytically tractable model of 1D interacting system, and, second, because it can be considered as a working paradigm for a large class of phenomena, where diffusion is constrained by confining geometries, which suppress the particles's transverse motion. As further proof of the periodic renewed theoretical interest on this simple model, very  recently SF dynamics has been  reintepreted within the context of the large fluctuation theory \cite{krapivsky2015,sadhu2015,sabhapandit2015,krapivsky2014,krapivsky2015dynamical}.

From a historical perspective, the theory of SF dynamics was established already by the early Seventies, thanks to the work of Levitt \cite{levitt1974}, who first derived the asymptotic law in Eq. (\ref{SFD}), and Percus \cite{Percus_PRA_1974}, who generalised it to 1D files of arbitrary stochastic dynamics, namely,

\begin{equation}
 \langle \left[x(t)-x(0)\right]^2\rangle=\frac{\langle\left|X(t)\right|\rangle}{\rho}.
  \label{SF_Percus}
\end{equation}

\begin{figure*}
\includegraphics[scale=0.35]{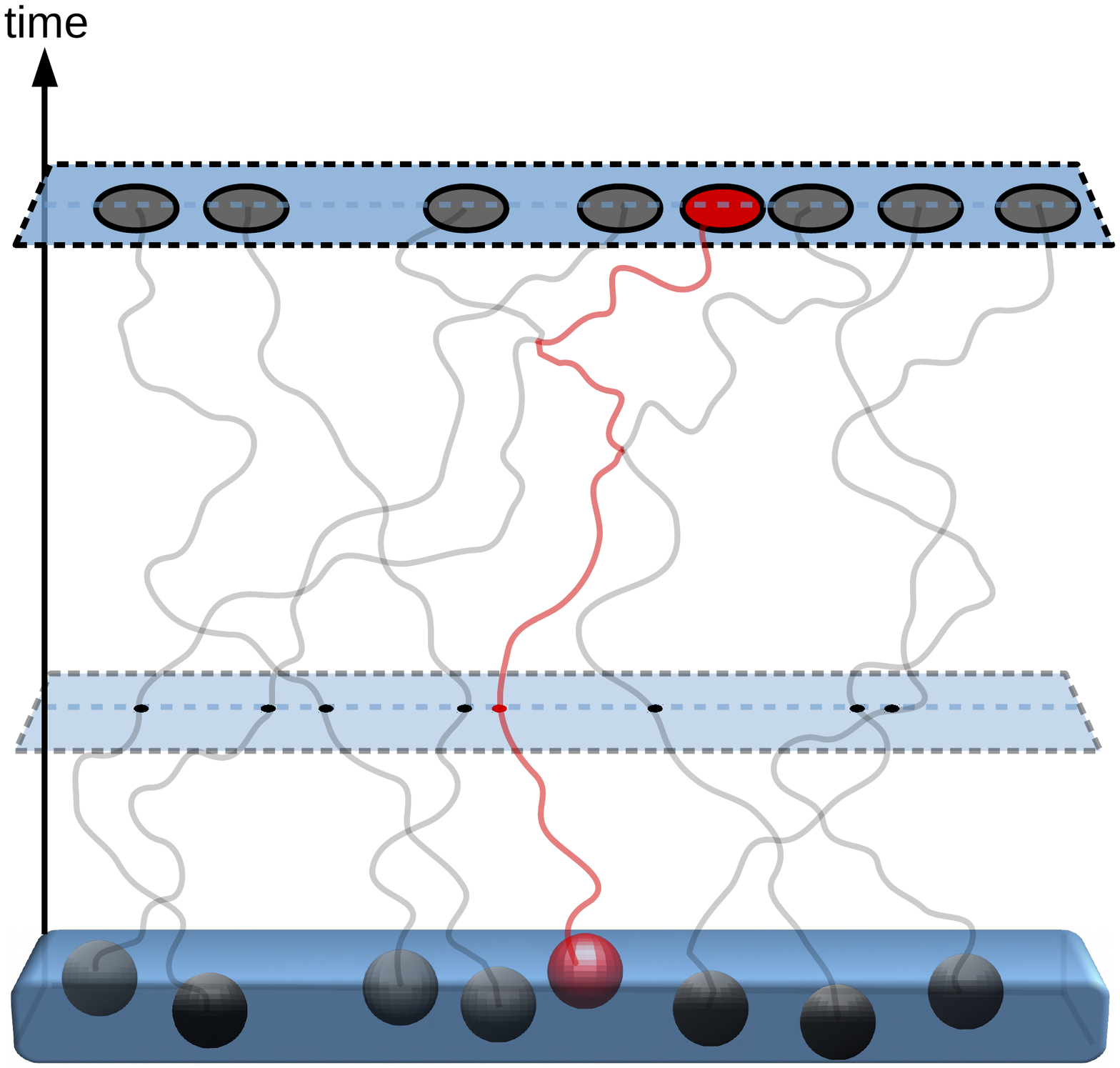}
\caption{\label{fig:1}{\bf Single file dynamics schematic representation.} A single file system is composed by Brownian particles diffusing within a narrow channel. The amplitude of this channel forbids the particle overtaking during the dynamics: the bottom part of the figure represents the assembly of equal mass diffusants in a quasi-one-dimensional squared channel. The Harris solution of SF diffusion schematises the particles as point-like and the corresponding stochastic trajectories as piece-wise curves evolving in time. Since particles are assumed to be zero-sized and the collisions purely elastic, the single file condition reduces to the relabelling of the particles number consequent to the requirement of maintaining the initial ordering (e.g. the fifth particle will always be the fifth in SF systems). Hence, a tagged particle trajectory (red particle and curve) is the composition of the \emph{free} particle trajectories, i.e., those Brownian paths which cross without interaction. The Percus formula (\ref{SF_Percus}) for the tagged particle asymptotic diffusion indeed, beautifully connects the SF dynamics to that of a non-interacting Brownian particles system. If the particle are assumed to have a finite size $\sigma$ (as in the upper part of the figure) the Percus' rule is modified just by replacing the average density $\rho$ with the excluded volume $\frac{1-\rho\sigma}{\rho}$.}
\end{figure*}

Here $\left|X(t)\right|$ denotes the absolute dispersion of a free diffusing particle from its initial position (see Fig.\ref{fig:1}). For instance, if the dynamics of the non-interacting particles is deterministic, $\left|X(t)\right|=Vt$, and the tracer asymptotic behaviour is diffusive with diffusion coefficient equals to ${\langle\left|V\right|\rangle}/{2\rho}$. Here the brackets must be intended as an average over a distribution of the velocity $V$, invariant under collisions in 1D and, therefore, constant in time \cite{lebowitz1968}. If, instead, the free dynamics is Brownian, $\langle\left|X(t)\right|\rangle=\sqrt{{4Dt}/{\pi}}$, and Eq. (\ref{SF_Percus}) reduces to Eq. (\ref{SFD}). In general, Eq.(\ref{SF_Percus}) can be employed to describe complex scenarios, where the tagged particle undergoes anomalous diffusion, either super- or sub-diffusive. Moreover,  finite-size corrections are taken into account  by replacing $\rho$ with the excluded volume term ${\rho}/({1-\rho \sigma})$, where $\sigma$ is the particle linear dimension. In this case, one talks of 1D rods gas (or Tonk's gas), the thermodynamics of which was fully explored by Tonk in 1936\cite{tonks1936}.

In the late Seventies, the problem of SFD was independently taken up by solid-state physicists, investigating superionic conduction. Richards\cite{richards1977} found by Monte Carlo simulations that for long times the tracer mean-square displacement (MSD) increased proportionally to $\sqrt{t}$; Fedders\cite{Fedders_PRB_1978} obtained a similar asymptotic behaviour by a diagrammatic Green's function technique; Alexander and Pincus \cite{Alexander_PRB_1978} were the first to connect the diffusive SF dynamics to the fluctuation of the collective particle density.

The popularity gained by the SF model in the last 30 years goes in hand with the increasing capability to detect and manipulate transport phenomena in both micro- and nano-environments. A typical example is represented by diffusion in zeolites, probably the richest playground for the study of anomalous transport in confined geometries. Zeolites are ubiquitous constituents of vugs and cavities of basalts and other traprock formations, which naturally form  channel networks that act as molecular sieves for molecules able to seep through. The vast literature accumulated in this field accounts for theoretical studies \cite{Karger_PRA_1992,rodenbeck1998,demontis2006,demontis2004,sastre1999,sastre2003,valdes2006,chou1999} and experimental realizations, providing positive evidence of the diffusive  regime predicted in Eq.(\ref{SFD}). A well-known example is represented by the diffusion of methane\cite{gupta1995}, $CF_4$ \cite{hahn1996} and  cyclopropane\cite{jobic1997,kukla1996} in  $AiPO_4-5$. Moreover SF diffusion has been found to be of fundamental importance in the understanding catalytic reactions such that  of cyclopentane on Pi/Mordenite. \cite{lei1993}

The tracer's subdiffusion law of Eq. (\ref{SFD}) has been demonstrated in a variety of laboratory experiments, involving paramagnetic particles in circular channels\cite{wei2000}, water suspended silica particles diffusing in 1D polydimethysiloxane channels \cite{lin2002,lin2005} and polystyrene particles diffusing in laser-created ring patterns \cite{lutz2004,lutzCond2004}, charged millimetric stainless steel balls moving in both straight and circular channels \cite{Coupier_EPL_2007,coste2014,delfau2012,delfau2012_bis,Delfau_PRE_2010}, and colloids in 1D energy random landscapes \cite{hanes2012}. Additionally, these studies contributed to shed light on how interactions between particles, both hydrodynamic, and magnetic or electrostatic, modify the SFD law of Eq. (1). They also helped elucidate the role of the confining potential in enhancing the tracer's (sub)diffusivity. Additionally, the escape process of N colloidal 
particles constrained in microfluidic channels has been shown to  be described in terms of the survival probability of the last particle to leave the channel, while its mean escape time of the process scales
inversely with D \cite{locatelli2016}.  Motivated by these findings, a remarkable theoretical effort focused on the study of SFD in colloidal systems\cite{herrera2007,herrera2008,herrera2010,euan2012,speranza2011,lizana2009,savel2006}, which ultimately led to the generalisation of the asymptotic law (\ref{SFD}) to assemblies of short-range  interacting Brownian particles\cite{Alexander_PRB_1978}. In parallel, SF theories developed to incorporate particles interaction with the confining walls\cite{Taloni_PRL_2006,Barkai_PRL_2009,gov2012,Ben-Naim_PRL_2009,hanes2012}, like, for instance, particles confined in compartmentalised narrow (asymmetric) channels. In this case, interaction potential maxima model the entropic barriers opposing the particle diffusion through the compartment pores\cite{burada_ChemPhysChem_2009}. Moreover, the widespread use of devices able to operate in nano- and microenvironments inspired a series of theoretical works where an external potential indirectly perturbs the entire file by acting on one particle only\cite{Taloni_PRE_2011,Taloni_PRE_2008,Leibovich_PRE_2014,lomholt2014}. Practical  examples are optical or magnetic tweezers that exert an external perturbation on the entire file by operating at the level of a single particle. This branch of research has led to the derivation of compact mathematical expressions for transport relations and fluctuation theorems, based on the (fractional) Langevin equation formalism\cite{Taloni_PRE_2008,Taloni_MNP_2013,Lizana_PRE_2010,Taloni_PRL_2010,Taloni_PRE_2011,taloni2014}.

A comprehensive overview of the widespread presence, detection and applications of  SF dynamics in material science,  must include carbon nanotubes, which recently have been shown to perform as high selectivity and throughput channels in synthetic membranes. Molecules inside a nanotube obey, indeed, the SF condition \cite{das2010,cambre2010,mukherjee2007,sholl2006}. SF trasport phenomena are observed also in  Wigner crystals driven by an external force on the surface of superfluid $_4He$ in the "quantum wire" regime \cite{vasylenko2014,klier2000}, electrons trapped on the surface of liquid helium moving in constriction geometries\cite{rees2014,klier2000}, polystyrene particles  in PDMS channels\cite{dettmer2014}, and Xe atoms inside the micropores of tris(o-phenylenedioxi)cyclophosphazene\cite{meersmann2000}.
Moreover, a system composed by colloidal particles confined in a channel whose width allows the particles overcoming, has been observed to undergo a SF-to-Fickian crossover just by varying the density of the particles within the channel \cite{siems2012}. Finally, we mention an interesting experimental study on the moisture expansion of several modern and ancient clay brick ceramics from the time of manufacture: the expansive strain increases as $(age)^{1/4}$ approximately, suggesting a possible new method for archaeological dating of ceramics\cite{wilson2003}.

Single file in biological processes is a relatively new field of investigation. For proteins, such as aquaporins AQP1, the selectivity for small polar solutes through the bylipidic membrane is well established\cite{hub2008}. Motor proteins, such as kinesin, dynein, and certain myosins, step unidirectionally along linear tracks, specifically microtubules and actin filaments, and play a crucial role in cellular transport processes, organization, and function \cite{kolomeisky2007}. Although the motion of a motor protein along filaments in principle involves a non-zero hopping probability  due to the motors ability  to move around the filament, or to a detachment-reattachment mechanism, a transient proteins jamming can be explained in terms of SF dynamics. Indeed,  Li  \emph{et al.} observed the typical SF dynamics while investigating the protein sliding along DNA \cite{li2009}. Moreover, elongated colloidal (rodlike) particles, such as chiral \emph{fd} viruses diffusing into liquidlike columnar hexagonal arrays, have been shown to undergo the typical collective motion of SF systems, by tracing the single-particle dynamics thanks to fluorescence labeling. Each particles have a finite probability to jump from one column to another, thus avoiding the SF condition and trespassing the neareast neighbours \cite{naderi2013}. In general, SF dynamics and transport properties are often observed and postulated to explain sublinear diffusivity of macromolecules in the cell's crowded environment\cite{milescu2005,auzmendi2012,marabelli2013,hofling2013,banks2005}.

On the theoretical side, SF has been shown to belong to the same universality class of Edward-Wilkinson\cite{bustingorry2010} and Rouse chains \cite{Lizana_PRE_2010,Taloni_PRL_2010}, thus extending its domain of applicability to simple polymer models. The first connection with polymer physics is probably hidden in the celebrated de Gennes' reptation paper \cite{de1976}, where it was shown that a defect (repton) diffuses within the strand nearly according to Eq.(\ref{SFD}). Remarkably, SF model has been successfully applied to recover the  conformational fluctuations of the donor-receptor distance within a protein \cite{murray2007}. The growing interest in SF processes in biology, stimulated a heterogeneous literature including first passage problems \cite{sabhapandit2007}, ergodicity problems \cite{Taloni_EPL_2012}, optimal  interaction between  channel and  particles \cite{zilman2009,berezhkovskii2005} and the dependence of the diffusion law, Eq.(\ref{SFD}), on the file initial conditions \cite{Leibovich_PRE_2014,Taloni_EPL_2012,Lizana2014}, the  distributed mass, the friction coefficients, the number of the particles in finite files \cite{lizana2009}, and the stochastic dynamics entering Eq.(\ref{SF_Percus}). As a matter of fact, when the free particles dynamics is non-Brownian but subdiffusive, the tagged particle dynamics is  affected giving rise to clustering and unexpected scenarios like order-disorder phase transitions\cite{flomenbom2011}. Correspondingly, in recent years there has been a considerable thrive  of models  partially modifying the strict SF condition: models with a finite passing probability\cite{lucena2012,kutner1984}, with inelastic collisions \cite{cecconi2004,savel2006} , and with special initial density profiles\cite{Flomenbom_EPL_2008,flomenbom2010,tripathi2014,flomenbom2014,aslangul1998}.

With this review article, we intend to offer an interdisciplinary introduction to SF processes, their realisations, and technological applications. Our discussion will be particularly focused on SF systems in soft matter, being this the area of research that contributed the most to the recent advances of this topic, both experimentally and theoretically.

\section{Formulation of SF dynamics and tagged particle anomalous diffusion}

The SF model is formally introduced as an assembly of interacting Brownian particles on a line. The interaction is primarily assumed to be hard-core, so that the particles retain their initial ordering while moving. This means that one can pick any
file particle and describe its stochastic dynamics; this
particle is usually referred to as the \emph{tagged particle} or tracer. Moreover, to simplify this task, in Harris' first formulation  momentum and energy were conserved at each collision \cite{Harris_JApplProb_1965}, so that the pair collisions cause a mere relabelling of the particles's indices. Taking advantage of this obvious symmetry,  different sophisticated  approaches have been put forward in the past to derive asymptotic diffusion and transport properties of the tagged particle (see Refs.\cite{Percus_PRA_1974,van1983,Karger_PRA_1992}, among others). We next review some of the most recent refinements of this approach.

\subsection{SF probability density function}

A SF system consisting of $2M+1$ particles is described by the positions and velocities of all its particles at time $t$, $\left(\mathbf{x},\mathbf{v};t\right)$, where $\mathbf{x}\equiv \left\{x_q\right\}$ and the particle index $q$ is restricted by $-M \leq q\leq M$. Hence, the probability density function (PDF) in the phase space is denoted by $P\left(\left.\mathbf{x},\mathbf{v};t\right|\mathbf{x^0},\mathbf{v^0};0\right)$. In the Smoluchowski approximation, one can drop the velocities and write the equation governing the PDF as a simple diffusion equation. In the case of
point-like particles, such equation reads

\begin{equation}
  \frac{\partial }{\partial t}P\left(\left.\mathbf{x};t\right|\mathbf{x^0};0\right)=D\sum_{q=-M}^{M}\nabla_q^2 P\left(\left.\mathbf{x};t\right|\mathbf{x^0};0\right),
  \label{SF_density_eq}
\end{equation}

\noindent where $D$ is the single particle diffusion coefficient and $\nabla_q\equiv \frac{\partial}{\partial x_q}$. The initial condition
for solving Eq.(\ref{SF_density_eq}) is $P\left(\left.\mathbf{x};0\right|\mathbf{x^0};0\right)=\prod_{q=-M}^M\delta\left(\mathbf{x}-\mathbf{x^0}\right)$  where the $\delta$ stands for the Dirac's delta distribution. By imposing reflecting boundary conditions (SF system in a box) and taking advantage of the label exchange which describes pair collisions, Eq.(\ref{SF_density_eq}) can be solved by having recourse to Bethe's ansatz \cite{kumar2008,lizana2009,Lizana_PRL_2008,flomenbom2014}. Thus, given the  multi-particle PDF $P\left(\left.\mathbf{x};t\right|\mathbf{x^0};0\right)$, one derives the reduced tracer's PDF, which is asymptotically Gaussian with variance equal to the MSD of Eq. (\ref{SFD}). Remarkably, this analysis has shown the existence of three different regimes undergone by the tagged particle MSD, allowing the analytical determination of any relevant time scale.   The exact solution provided for the PDF (\ref{SF_density_eq})  is particularly suitable to describe realistic situations where the file is confined within pores with open or closed boundaries, such those reproduced in the experiments of  Ref.\cite{locatelli2015,locatelli2016}. This analytical framework can be easily generalized to the more realistic cases of files with distributed  diffusion coefficients    \cite{Ambjornsson2008,flomenbom2010}, confined to quasi-1D domains \cite{burada_ChemPhysChem_2009}, or diffusing in asymmetric  landscapes \cite{jara2006}.

\subsection{Microscopic dynamics of SF systems}

A different and versatile point of view consists in deriving macroscopic observables starting from the stochastic dynamics of the particles in the files. According to this viewpoint, one can write a set of Langevin equations (LE) for the entire system,

\begin{equation}\frac{d^2x_q(t)}{dt^2}=-\gamma
\frac{dx_q(t)}{dt}-\nabla_qU\left(\vec{x};X,t\right)+\eta_q(t),
\label{LE}
\end{equation}

\noindent where the damping constant $\gamma$ and the
noises $\eta_q(t)$ satisfy the fluctuation-dissipation relations,

\begin{equation}
\langle\eta_q(t)\rangle=0,
\label{FD0}
\end{equation}

\begin{equation}
\langle\eta_q(t)\eta_l(t')\rangle=2k_BT\gamma\delta_{q,l}\delta(t-t')
\label{FD}.
\end{equation}

\noindent
The potential function in Eq. (\ref{LE}) consists of two
contributions,

\begin{eqnarray} & & U\left(x_{-M},\dots,x_M;X,t\right) \\ \nonumber & &=\sum_{q\neq l=-M}^{M}
U_{HC}\left(\left|x_q-x_l\right|\right)+U_{\rm
int}\left(x_{-M},\dots,x_M;X,t\right), \label{potential} \end{eqnarray}

\noindent with the hard-core repulsion,

\begin{equation}
U_{HC}\left(\left|x_q-x_l\right|\right)=
\left\{
\begin{array}{ccc}
\infty  & & |x_q-x_l|=0\\
0       & & {\rm otherwise},
\end{array}
\right.
\label{potential_HC}
\end{equation}

\noindent ensuring the SF condition. Note that the spatial
coordinate, $X$, has been capitalised so as to be distinguished from
the particle trajectories, $x_{-M}(t), \dots, x_M(t)$. The additional
substrate potential $U_{\rm int}$, is often introduced to model
distinct physical situations, such as the interaction with a substrate (landscape potential)\cite{Taloni_PRL_2006,Barkai_PRL_2009,Barkai_PRE_2010,Ben-Naim_PRL_2009,lucena2012}, single particle force field~\cite{Taloni_PRE_2011,Burlatsky_PRE_1996, Taloni_PRE_2008,Taloni_MNP_2013,Leibovich_PRE_2014,Villamaina_JSTAT_2008}, coupling of both, and interparticle interactions \cite{taloni2014}.
The latter  is particularly suited to describe real physical situations (with particular emphasis on colloidal suspension), where the SF particles are taken to be electrically or magnetically charged. Experimental realisations have been carried out using particles with magnetic
~\cite{Wei_Science_2000}, electric dipole \cite{lin2002,lutzCond2004,lutz2004}, or screened electrostatic
pair interactions \cite{Coupier_PRE_2006, Coupier_EPL_2007,Coste_PRE_2010, Delfau_PRE_2010}. Correspondingly, a considerable effort has been devoted to the detailed  modeling of these systems and their numerical implementation \cite{euan2012,herrera2010,nelissen2007,savel2006}.

\noindent Given the set of LE (\ref{LE}) for the entire system, the question is whether it is possible to derive an effective tracer's LE, where the interactions with the neighbouring particles can be incorporated in a single noise term. This issue was solved in Ref.\cite{Taloni_PRE_2008} where a generalised LE was shown to reproduce quite accurately the tagged particle dynamics, both in underdamped and overdamped limits. However, a rigorous derivation for the overdamped limit was obtained only in Ref.\cite{Lizana_PRE_2010}. In this paper indeed, it was shown that a stochastic equation governing the tracer motion can be derived from the  LE (\ref{LE}) through a procedure called \emph{harmonization}. Such an equation is a fractional Langevin equation (FLE) and takes the following form,

\begin{equation}
2\sqrt{\gamma\,k_BT}\rho\frac{d^{1/2}x(t)}{dt^{1/2}}
=\xi(t), \label{FLE}
\end{equation}

\noindent where we dropped the index $q$ to simplfy the notation. The Caputo fractional derivative is defined by
\cite{Samko,Podlubny}

\begin{equation}
\frac{d^{1/2}f(t)}{dt^{1/2}}=\frac{1}{\Gamma\left(\frac{1}{2}\right)}\int_0^t\frac{df(t')/dt'}{\left|t-t'\right|^{\frac{1}{2}}}\,dt',
\label{caputo_der}
\end{equation}

\noindent and the (fractional) noise $\xi_q(t)$ is assumed to be Gaussian, zero-mean valued and related to the damping kernel by Kubo's generalised
fluctuation-dissipation relation \cite{Kubo_1966}. This formalism offers
the advantage that all observables of practical
interest can be calculated analytically. For instance, it turns out
that the subdiffusive law of Eq.(\ref{SFD}) is
closely related to the persistent memory effects that result in the
negative power-law tails of the velocity autocorrelation functions
\cite{Percus_PRA_1974,VanBeijeren_PRB_1983, Marchesoni_PRL_2006,Taloni_PRE_2006, Felderhof_JCP_2009,Tripathi_PRE_2010,cecconi2003}.  This demonstrates  that the tracer anomalous diffusion is a mere consequence of the non-Markovian nature of its stochastic dynamics and of the underlying collisional process \cite{Marchesoni_PRL_2006,Taloni_PRE_2006}. Moreover, in the long time limit the two-time correlation function is $\langle \left[ x(t)-x(0)\right]\left[x(t')-x(0)\right]\rangle \sim \sqrt{t}+\sqrt{t'}-\sqrt{|t-t'|}$. This is precisely the definition of a  fractional Brownian motion \cite{Mandelbrot_1968} with an Hurst exponent $H=1/4$. This finding has considerably broaden the theoretical relevance of single-file models,  clarifying the intimate connection between single-file and other linearly interacting many-body systems \cite{Taloni_PRL_2010}. As a matter of fact, the SF model shares with a wide class of stochastic models the remarkable propriety that any correlation function attains a universal scaling form, which can be uniquely expressed in terms of compact scaling relations, and that can in turn lead to the anomalous behaviour of physical observables. This class of systems goes under the name of generalised elastic model (GEM) \cite{Taloni_PRL_2010,Taloni_EPL_2012,Taloni_MNP_2013,Taloni_PRE_2011,taloni2016}. Also among GEM's the SF model is regarded as a paradigm, due to the simplicity of its formulation and the few parameters needed to define the underlying microscopic dynamics. The FLE (\ref{FLE}) has been validated by
extensive numerical simulations \cite{Taloni_PRE_2008,Lizana_PRE_2010} and proved to hold also in the presence of a field of force acting on a single file particle\cite{Lizana_PRE_2010,Taloni_PRE_2011,taloni2016,taloni2014}.

Remarkably, the FLE can be derived from a diffusion-noise approach \cite{Taloni_PRE_2008, taloni2014}.  Indeed, starting from the Langevin equations (\ref{LE}) and considering
noninteracting SF's with $U_{\rm int}=0$, it is possible to write down  the stochastic equation governing the time evolution of the collective density  profile along the 1D substrate. In doing so, one formally links the density fluctuations to the particle stochastic dynamics, a general result that incorporates and extends the well-known Alexander-Pincus relation \cite{Alexander_PRB_1978}.

\subsection{The mobility factor}

By rewriting Eq. (\ref{SFD}) as

\begin{equation}
  \langle \left[x(t)-x(0)\right]^2\rangle=2F\sqrt{t},
\label{SFD_mobility}
\end{equation}

\noindent one introduces the so-called mobility coefficient\cite{Karger_PRA_1992} $F=\sqrt{\frac{D}{\pi\rho^2}}$  or\cite{Levitt_PRA_1973,Percus_PRA_1974}

\begin{equation}
  F=\frac{1-\rho\sigma}{\rho}\sqrt{\frac{D}{\pi}},
\label{SFD_mobility_true}
\end{equation}

\noindent when the particles have linear size $\sigma$. For a large class of overdamped systems and finite-range interaction potentials, a situation often fulfilled by colloidal suspensions, Kollmann extended Alexander and Pincus result \cite{Alexander_PRB_1978} showing that the mobility factor in the Fourier $q$-domain takes the form,

\begin{equation}
  F(q)=S(q,t=0)\sqrt{\frac{D_c(q)}{\pi\rho^2}},
\label{mobility_kollman}
\end{equation}

\noindent under the condition that $q\ll\frac {2\pi}{\sigma}$. The functions $S(q,t=0)$ and $D_c(q)$ are, respectively, the static structure
factor and the short-time collective-diffusion coefficient in the Fourier domain. From Eq. (\ref{mobility_kollman}), the long-time character of SFD is determined by the short-time collective
dynamics at long wavelengths. In such a limit, $S(q,t=0)$ tends to $S(0,t=0)$, where $S(0,t=0)$ in monodisperse systems corresponds to the normalised isothermal compressibility \cite{nagele1996}. The latter quantity can be measured  both in experiments\cite{lutz2004,lin2002,lin2005} and computer simulations \cite{herrera2010}. By expressing $D_c(q)$ as $D_c(q)= D{H(q)}/{S(q,t=0)}$, where $H(q)$ is the hydrodynamic factor, in the limit $q \to 0$ Eq.(\ref{mobility_kollman}) can be rewritten as,

\begin{equation}
  F(q)=\sqrt{\frac{DS(q,t=0)H(q)}{\pi\rho^2}}.
\label{mobility_kollman_hydro}
\end{equation}

\noindent This equation incorporates the effects of the short-time single particle diffusion (in $D$), density fluctuations (in $S(q)$), and  hydrodynamic interactions (in $H(q)$). When hydrodynamic interactions are negligible, as in the case of the experiment reported in \cite{lin2005}, $H(q)$ is set to 1 and Eq.(\ref{mobility_kollman_hydro}) reduces to the more familiar Eq. (\ref{SFD_mobility_true}), after using for the isothermal compressibility the expression valid for Tonk's gas \cite{tonks1936}, i.e.,  $S(q=0;t=0)=(1-\rho\sigma)^2$.

\noindent Making  use of sophisticated numerical techniques, San\'e et al.\cite{sane2010} corroborated the long-time correctness of the relation of Eq. (\ref{SFD_mobility_true}) with the explicit inclusion of the hydrodynamic interactions. Recently, Herrera-Velarde et al.\cite{herrera2010} explored systematically both the structural and the dynamical properties of three different interacting files, namely, WCA (Weeks–Chandler–Andersen), Yukawa and superparamagnetic particles in a 1D channel (in the last case the interaction potential $U_{\rm int}\left(x_{-M},\dots,x_M;X,t\right)$  in \ref{potential} reduces to a pairwise potential $U_{\rm int}\left(x_{-M},\dots,x_M\right)=\sum_{q\neq l=-M}^{M}\frac{1}{\left|x_q-x_l\right|^3}$).  When the system is highly structured and the particles spatially correlated at long distances, the file dynamics gets dramatically suppressed and characterized by a unique mobility factor. These characteristics suggest the occurrence of a structural transition from a fully disordered to a pseudo-solid state in repulsive 1D files. Such a transition seems to take place when the main peak of the structure factor grows larger than $S(q_{max},t=0)\simeq 7$; this condition corresponds to values of the reduced mobility factor of the order of $\frac{F\rho}{\sqrt{D}(1-\rho\sigma)} \simeq 0.1$. In this regard, $F$ would play the role of the control parameter of the advocated order-disorder-like transition.

\section{SFD scaling relations}

The asymptotic law of Eq. (\ref{SF_Percus}) connects tracer's diffusion with the absolute dispersion of a free particle.
In Ref.\cite{Flomenbom_EPL_2008} a general scaling relation  between the mean absolute displacement of the tagged and a free particle was derived:

\begin{equation}
  \langle \left|x(t)-x(0)\right|\rangle\sim\frac{\langle \left|X(t)\right|\rangle}{n},
  \label{SFD_flom}
\end{equation}

\noindent where $n$ is the number of particles in the length $\langle \left|x(t)-x(0)\right|\rangle$. When the density profile is uniform in average $n=\rho\langle \left|x(t)-x(0)\right|\rangle$ and one recovers Percus' relation (\ref{SF_Percus}). The number $n$
is said \emph{cooperation term}, since all $n$ particles must cooperate and move in the same direction for the tagged particle to finally reach the endpoints of the interval $\langle \left|x(t)-x(0)\right|\rangle$.

\subsection{Heterogeneous files}

The scaling Eq. (\ref{SFD_flom}) has the valuable property of applying also to situations when the particle density along the 1D substrate is  nonuniform at $t=0$. Let us consider, for instance, a file with density decaying with the distance from the origin, $X$, like \cite{flomenbom2010}

\begin{equation}
  \rho(X)\sim \rho_0\left(\frac{X}{\Delta}\right)^{-\alpha},
  \label{rho_pl}
\end{equation}

\noindent  with $0\leq \alpha\leq 1$ and $X >\Delta$. Taking advantage of the relation (\ref{SFD_flom}), one expresses the tracer's absolute dispersion as $\langle \left|x(t)-x(0)\right|\rangle\sim \langle \left|X(t)\right|\rangle^{\frac{1+\alpha}{2}}$.  Correspondingly, the tracer MSD scales like $\langle \left[x(t)-x(0)\right]^2\rangle\sim t^{\frac{1+\alpha}{2}}$ and its PDF is Gaussian function \cite{Flomenbom_EPL_2008}. These results highlight a smooth interpolation between the SF and the corresponding free-particle dynamics, a property associated with the spatial expansion of the file: from the denser area in the middle, the file particles move towards the more diluted periphery. Such a mechanism can be interpreted as driven by an effective force (pressure) pushing outwards. This finding helps elucidate how local density fluctuations may affect the tracer's dynamics: the particles move faster than a typical SF when density fluctuations are higher than in Eq. (\ref{rho_pl}) and, vice versa, move slower when fluctuations are smaller. Aslangul\cite{aslangul1998} studied a related process by considering a file with $2M$ particles all located
 around the origin at $t=0$. In his analysis, the edge particles diffuse like free particles, and the middle particles behave like regular SF particles. Expanding files were also studied in Ref.\cite{tripathi2014}.

Another interesting case where Eq. (\ref{SFD_flom}) can be of valuable use is when  the particles’ diffusion coefficients are heterogeneous. The diffusion distribution is assumed to be a power-law with exponent $\gamma$:

\begin{equation}
  W(D)=\frac{1-\gamma}{\Lambda} \left(\frac{D}{\Lambda}\right)^{-\gamma},
  \label{D_pl}
\end{equation}

\noindent  where $0\leq \gamma <1$. A rigorous treatments shows, indeed, that the $W(D)$ of Eq.(\ref{D_pl})
is a limiting distribution, in the sense that for any other distribution with finite first moment, one recovers the familiar SF diffusion law of Eq. (\ref{SFD}) with effective diffusion coefficient $\langle D\rangle =\int dD W(D) D$ \cite{jara2006,Ambjornsson2008,lomholt2014}. When the diffusion coefficients are distributed according to Eq.(\ref{D_pl}) and the density profile is modeled by a power-law like Eq.(\ref{rho_pl}), one talks of \emph{heterogeneous files}.  For this broad a class of SF's, Eq. (\ref{SFD_flom}) yields immediately the generalized diffusion relations  $\langle \left|x(t)-x(0)\right|\rangle\sim \langle \left|X(t)\right|\rangle^{\frac{1-\gamma}{2c-\gamma}}$, with $c=\frac{1}{1+\alpha}$, and $\langle \left[x(t)-x(0)\right]^2\rangle\sim t^{\frac{1-\gamma}{2c-\gamma}}$.

Moreover, the concept of heterogeneous files can be further generalized to files of particles that differ for mass, size, or composition, and densities governed by the environment, rather than the mere file dynamics. For instance,  diffusive processes in living cells may extend over denser and less dense areas, as the result of the biological activity in the cell.

\subsection{Slow renewal files}

In Ref. \cite{flomenbom2010renewal} was  shown that the only requirement for the relations (\ref{SF_Percus}) and (\ref{SFD_flom}) to be valid is that the single particle stochastic dynamics is renewal, i.e.  a  process  where all particles  move at the same time. For example, taking the free stochastic dynamics to be a subdiffusive process regulated by a continuous time random walk model (see \cite{metzler2000} and references therein), e.g. taking the distribution function of waiting times $\psi(\tau)\sim \tau^{-(1+\epsilon)}$ ($0<\epsilon<1$), therefore $\langle \left[x(t)-x(0)\right]^2\rangle\sim t^{\epsilon/2}$ \cite{flomenbom2010renewal,flomenbom2010,Flomenbom_EPL_2008,flomenbom2014,flomenbom2010renewal,Bandyopadhyay_EPL_2008}.
Results involving renewal anomalous files are of value in all those cases where
the channel changes states, with times drawn from $\psi(\tau)$, from a state where the particles diffuse, to a state where the particles are bound to the channel quickly. These include: pores under on-off fields or under temperature changes, sensing devices (as it was suggested for zeolites \cite{valdes2006}) under on-off fields, channels as DNA sequencing devices under on-off fields \cite{schmalzing1998}.

\subsection{Clustering}

When the stochastic dynamics of each particle is different, i.e. particles jump at different times according to the distribution $\psi(\tau)$, the tracer MSD slows down considerably

\begin{equation}
  \langle\left[x(t)-x(0)\right]^2\rangle\sim\ln^2(t).
  \label{MSD_cluster}
\end{equation}

\noindent Moreover, after a transient time the particles form clusters
in such files, defining a dynamical phase transition: a large percentage $\xi$ of the particles are trapped in  clusters in the limit of small $\epsilon$, this percentage scales as a function of $\epsilon$ as $\xi\sim \sqrt{1-\epsilon^3}$.  Since clustering occurs only for anomalous $\epsilon$, $\xi$ describes the criticality of a phase transition. Indeed, it has a typical form for a scaling function in critical phenomena. The clustering phenomenon is observed only for long-tailed distribution $\psi(\tau)$, i.e. for $0<\epsilon<1$, while for distributions allowing the existence of a well-defined characteristic time, say exponential, for instance, the usual SF relation \ref{SFD} holds. Clustering is expected to be universal, since $\epsilon$
 is the only control parameter. Therefore, clustering might be applied in
 regulating biological channels, an important theme in biophysics \cite{chung2007}: by tuning $\epsilon$, in principle, it would be possible to pass from a clogged file  ($\epsilon<1$) to  a fluid file ($\epsilon>1$). When controlling the synchronisation of the particles, either having a file made of synchronised anomalous particles or having a file made of independent
anomalous particles, we have the possibility of seeing either a diffusing phase (for synchronised particles) or a clustered phase (for independent particles).

\section{SFD in colloidal systems}

As summarised in the previous sections, during the last decade, the influence of interactions on SFD of colloidal suspensions has been studied extensively\cite{prestipino2005,prestipino2013,Taloni_PRL_2006,herrera2007,herrera2008,herrera2010,euan2012,speranza2011,lin2002,lin2005,vasylenko2014,rees2014,Coupier_PRE_2006,delfau2012,delfau2012_bis,coste2014,Delfau_PRE_2010,Coste_PRE_2010,castro2014,zahn1997,rinn1999,franosch2011,meiners1999,reichert2004,fitzgerald2001,ryabov2011,euan2014}. In particular, recent experiments with charged
macroscopic particles (millimetric steel balls) confined in circular channels \cite{Coupier_PRE_2006} or
linear channels of finite length\cite{Delfau_PRE_2010,Coste_PRE_2010} exhibited particle diffusion slower than the scaling. Delfau et al.\cite{Delfau_PRE_2010} experimentally identified three different dynamical regimes, anticipated mathematically,\cite{coste2014,delfau2012,delfau2012_bis}, and found that the particle response to thermal fluctuations strongly depends either on the particle position in the channel or the
local potential it experiences. The slower diffusion found in the previous experiments can be explained in terms of the inertial (ballistic) dynamics of the steel balls\cite{Coste_PRE_2010}. When the particles diffuse in
a circle or a closed box, a geometrical time scale determines the saturation at very longtimes \cite{lizana2009,castro2014}. In a related process, that is, electrons diffusion on liquid helium, investigators reported unique behaviours, upon changing the width of the channel \cite{vasylenko2014,rees2014}. There,  the oscillations of the average particle velocity in channels with short constrictions exhibit a clear correlation with the
transitions between states with different numbers of rows of particles in the constriction, while for channels with longer constrictions these oscillations are suppressed. Again, this effect can be explained with the interactions of classical electrons among themselves and with the helium.

\subsection{The role of hydrodynamic interactions}

The effects of hydrodynamics interactions on the SFD have been barely studied \cite{sane2010}. Furthermore, the coupling of external fields together with the effects of the hydrodynamic interactions gives rise to a richer dynamical scenario, since the former breaks the homogeneous mass distribution along the file, whereas the latter induce important effects at long times  \cite{franosch2011,meiners1999,reichert2004,fitzgerald2001}. For example, it has been recently observed that the motion of a colloidal particle in a strong optical trap reveals (hydrodynamic) resonances at short-time scales in contrast to typical overdamped colloidal systems \cite{franosch2011}. However, in dilute colloidal dispersions, the effects of hydrodynamic interactions are small and can be simply ignored \cite{ryabov2011,hanes2012}. Nevertheless, there are other cases, e.g., at finite particle concentration, where the aforementioned coupling results in interesting dynamical modes at long times. In particular, Euan Diaz et al.\cite{euan2012} explicitly studied the hydrodynamic effects on the SFD in interacting colloidal systems subjected to a periodic external potential. These authors considered both weak and strong couplings and different values of the external potential strength, and found that hydrodynamic interactions enhance the particle mobility. In particular, it was observed that at long times, the tagged particle MSD scales like $\sim t^\mu$, with $0.5<\mu< 1$. In addition, it was shown that in files without external fields, the MSD deviates from the scaling law of Eq. (\ref{SFD}), being $\mu\simeq 0.56$. On the other hand, when the external potential is switched on, it is seen that particles are sensitive to the strength and commensurability of the external potential with the inter-particle spacing $\frac{1}{\rho}$. However, most of the dynamical modes observed by the authors can be explained in terms of collective diffusion, due to the long-range nature of the hydrodynamic interactions, and the competition between particle-particle and particle-substrate interactions. The latter ones are responsible for the  particles settling at the minima of the external potential. Nonetheless, in the particular case of $\frac{1}{\rho}= 1$, the long-time behaviour of the colloidal file exhibited three distinct dynamical regimes, which still need to be explored in detail to appreciate the anti-cooperative action of the hydrodynamic interactions.

\subsection{On the importance of the interaction potential}

By now, SF diffusion in systems made of interacting particles
with strongly repulsive and radially symmetric pair potentials is well-understood \cite{prestipino2005}. In this regard, the particular form of the interaction potential matters and is crucial for the understanding of the diffusion mechanisms in narrow channels. Typical examples are pair potentials with a soft-core, or a combination of short-rang attraction and long-range repulsion, or modeling highly-directional interactions. The simplest potential model for a soft-core fluid is the Gaussian-core model (GCM). Since the degree of softness in such a fluid depends on the temperature, its phase behaviour is completely different from that observed in fluids with a well-defined hard-core\cite{prestipino2005}. In particular, it has been found that in 1D, a GCM fluid exhibits thermodynamic and structural
anomalies \cite{prestipino2013,speranza2011}. A GCM fluid confined in a channel shows, at long-times, a crossover
from normal  to SF diffusion \cite{herrera2016}, such a transition depending the file's thermodynamic state\cite{herrera2010,herrera2016}.

More recently, the soft matter community has addressed SALR potentials, that is potentials with an excluded volume, short-range attraction (SA) and long-range repulsion (LR). Such complex potentials
are presently the focus of academic research, as well as industrial technology, for instance, of new materials. Therefore, a major effort is currently underway to explore the rich phase space due to the interplay of the competing short- and long-range interactions modeled by this class of potentials.
SALR fluids are of particular interest, as they can produce equilibrium states containing stationary clusters \cite{godfrin2014}. However, their dynamical behaviour under strong confinement has not been fully investigated, yet. A few theoretical studies advocate the formation of finite-size chains with a well-defined length distribution, which populate the channel landscape in the long time regime\cite{edith2016}.

It is worthy to mention that an increasing experimental effort has been devoted to the synthesis of complex colloidal particles with chemically or physically patterned surfaces and possible specific shapes far from spherical. These new colloidal particles with
highly-directional or anisotropic interactions are commonly named patchy particles, see, e.g., \cite{bianchi2011} and references therein. The anisotropy of the interaction and the limited valence in bonding are the salient features determining the collective behaviour of such systems. By tuning the number, the interaction parameters and the local arrangements of the patches, it is possible to investigate a wide range of physical phenomena, from different self-assembly processes of proteins, polymers and patchy colloids to the dynamical arrest of gel-like structures \cite{bianchi2011}. Nonetheless, the effect of the anisotropic interactions on SFD of patchy colloids remains to be explored.

\section{Looking ahead}

 Historically, the theory of single file diffusion has been explored within diverse contexts. The large heterogeneity of such approaches certainly highlighted the importance of the SF model as a benchmark in different area of physics, but on the other hand it has clarified the enormous wherewithals that  this relatively simple model may have  on technological applications. As a matter of fact, the advent of our ability to observe and manipulate systems in real time at the micro- and nanoscale has changed our perspective profoundly. Nowadays, any theoretically interesting correlation function by instance, is experimentally accessible. Some of the applications of this technological advance  are clear and immediate.

  \emph{Zeolites--} The system of channels and cavities is different in each zeolites structure, as are the effective sizes of entries parts, giving rise to a wide variety of materials, each capable of screening molecules and cations by molecular and ions sieving in slightly different manners. In recent years zeolite minerals have found increasing applications in the field of pollution abatement and they are fast becoming standard materials in the components of such facilities. They have been largely utilized in nuclear factories as radioactive waste-disposal, in the removal of $S0_2$ and other pollutants from stack gases of oil, or in the production of inexpensive, oxygen-enriched streams of variable degree of purity. Furthermore zeolites are employed as efficient heat exchanger in solar energy installations and they offer excellent prospects for the fabrication of photonic and optical materials.
  
  \emph{Carbon nanotubes--} Transport of water molecules and other liquids through single-wall carbon nanotubes (SWNTs) is a very active
  area of  research, in part because of the great potential for applications in nanofluidics and ultraselective molecular filtration.  This offers interesting opportunities for the fabrication of nanoelectro-fluidic devices such as nanopumps or hydroelectric power converters. Water filling of very thin SWNTs is also very appealing from a fundamental point of view, as this would create the ideal conditions to achieve a ``first-in-first-out'' single-file  transport regime, even over long distances and time scales.

  \emph{Micro- and nanofluidics--} Micro- and nanofluidic devices for colloidal particles   constitute a true playground for future innovations. Indeed one could control particle transport in technological applications such as, e.g., the construction of drift ratchets for particle sorting, or the design of optimized devices for  diffusion enhancement, assisted by an external force gradient. By  modulation of hydrodynamic interactions, channel geometry and file density instead, one can actually control normal-to-subdiffusive crossover, thus lowering or increasing the efficiency of nanoelectric motors.

  \emph{Biodevices--} The extreme selectivity at the molecular scale addressed previously is reminiscent of the selective transport of molecules through cell membranes via porine proteins. One could then imagine synthetic membranes with pores  filtering only some kind of proteins, which might be relevant for developing in vitro tests for early diagnosis of  diseases. Planar nanopores for DNA sequencing are designed based on SF first-passage processes.   Moreover the suppression of longitudinal diffusion on SF systems, is of prominent utility in  drug-releases devices.


\footnotesize{
\bibliography{rsc} 
\bibliographystyle{rsc} 
}

\end{document}